\documentclass{emulateapj}
\def\stacksymbols #1#2#3#4{\def\theguybelow{#2}
        \def\verticalposition{\lower#3pt}
        \def\spacingwithinsymbol{\baselineskip0pt\lineskip#4pt}
        \mathrel{\mathpalette\intermediary#1}}
\def\intermediary #1#2{\verticalposition\vbox{\spacingwithinsymbol
        \everycr={}\tabskip0pt
        \halign{$\mathsurround0pt#1\hfil##\hfil$\crcr#2\crcr
                \theguybelow\crcr}}}
\def\lta{\stacksymbols{<}{\sim}{2.5}{.2}}
\def\gta{\stacksymbols{>}{\sim}{3}{.5}}

\usepackage{float}

\shorttitle{Circumgalactic Oxygen Absorption and Feedback}
\shortauthors{Mathews \& Prochaska}
 
\begin{document}

\title{Circumgalactic Oxygen Absorption and Feedback}

\author{
William G. Mathews\altaffilmark{1} and 
J. Xavier Prochaska\altaffilmark{1} 
}

\altaffiltext{1}{University of California Observatories/Lick
  Observatory,
Department of Astronomy and Astrophysics,
University of California, Santa Cruz, CA 95064
(mathews@ucolick.org).}

\begin{abstract}
OVI absorption in quasar spectra caused 
by intervening circumgalactic atmospheres 
suggests a downturn in the atmospheric column density 
in sightlines passing beyond 
about 100 kpc from central star-forming galaxies.
This turnover supports the hypothesis that 
the oxygen originates in the central galaxies.
When converted into oxygen space density using
an Abel integral inversion,
the OVI columns require $\gta 10^9$ $M_{\odot}$ 
of oxygen concentrated near 100 kpc.
Circumgalactic gas within this radius 
cools in less than 1 Gyr and radiates  
$\sim 10^{42.2}$ erg s$^{-1}$ overall.
The feedback power necessary
to maintain such oxygen-rich atmospheres 
for many Gyrs
cannot be easily supplied by galactic supernovae.
However, massive central black holes
in star-forming galaxies
may generate sufficient
accretion power and intermittent 
shock waves at $r \sim 100$ kpc to balance 
circumgalactic radiation losses 
in late-type $L^{\star}$ galaxies.
The relative absence of OVI absorption observed 
in early-type,
passive $L^{\star}$ galaxies may arise from 
enhanced AGN feedback from their
more massive central black holes.
\end{abstract}


\section{Introduction}


The possibility of high oxygen abundances in hot, virialized
gas around normal star-forming galaxies has attracted 
much attention. 
Tumlinson et al. (2011)
describe strong UV absorption in quasar spectra  
due to the OVI doublet (1031.9, 1037.6\AA) 
having redshifts similar 
to those of $\sim L^{\star}$ galaxies 
observed close to quasar sightlines.
Distances between sightlines and the central galaxies,
i.e. the impact parameters $R$, 
are large, implying that the absorption 
occurs in extended circumgalactic gaseous atmospheres
surrounding star-forming galaxies. 
Impact parameters of $10 \lta R \lta 150$ kpc 
are observed in star-forming galaxies 
with redshifts $0.1 \lta z_{gal} \lta 0.4$ 
and stellar masses 
$9.5 \lta {\rm log}(M_{\star}/M_{\odot}) \lta 11.5$.
Large oxygen column
densities N$_{\rm OVI} \approx 10^{14.5 \pm 0.25}$ cm$^{-2}$ 
suggest large oxygen masses.

Detailed observations of circumgalactic absorption
by O$^{+5}$ and other ions of C, N, O, Mg, Si, and Fe are described 
in a series of publications: 
Werk et al. (2012,2013,2014,2016), 
Johnson, Chen, and Mulchaey (2015 $\equiv$ JCM), 
Borthakur et al. (2016).
Broad OVI absorption is detected in essentially    
all sightlines near star-forming galaxies, 
but in only a small fraction of sightlines near 
passive, early-type $L^{\star}$ galaxies. 
Of interest here are decreasing \ion{O}{6}
absorption columns that extend beyond the 
$\sim150$ kpc survey limit of 
Tumlinson et al.
to at least 
$\sim 300$\,kpc around $L^{\star}$ galaxies 
(Prochaska et al. 2011; JCM).

Meanwhile, a significant computational effort is 
underway by cosmological simulators to understand 
the origin of  
oxygen in the hot circumgalactic gas:
Stinson et al. (2013);
Suresh, et al. (2015); Liang, Kravtsov \& Agertz (2016);
Roca-Fabrega et al. (2016);
Oppenheimer, et al. (2016); Sokolowska et al. (2016) etc.
In these studies most of the 
circumgalactic oxygen is provided by supernova-driven 
galactic winds from the central galaxy. 
Nevertheless, many of these sophisticated 
computational studies underpredict 
observed OVI column densities. 

Recent theoretical treatments of \ion{O}{6} in the 
circumgalactic medium 
include the phenomenological model of 
Stern et al. (2016)
which envisions \ion{O}{6} as low-density, 
externally photoionized gas 
surrounding denser ionized gaseous clumps 
that produce lower ionization absorption lines of 
C, N, O, Mg, Si, and Fe.
This scenario explicitly links the kinematics of the 
\ion{O}{6} gas with the lower ionization material
(as suggested by observations of Werk et al. 2016).
Most recently, 
Faerman et al. (2016)
introduce a 
model for galaxy coronae to reproduce \ion{O}{6}
observations together with X-ray observations of higher
ionization states of oxygen in the Milky Way.  
In their phenomenological
model, circumgalactic gas is multi-phase with 
a distribution of densities 
all in pressure equilibrium. 

As with lower ionization absorption lines,
some OVI absorption may occur in small, 
actively cooling regions.
However, as seen in Figure 3 of Werk et al. (2013),
broad velocity widths of OVI absorption lines 
extend to velocities unassociated with
low ionization absorption.
This suggests that most OVI absorption arises not in 
thermally perturbed regions 
but in extended hydrostatic atmospheres -- 
we adopt this assumption here.

\begin{figure*}[ht]
\begin{center}
\vspace{-0.3cm}
\includegraphics[width=6.5in]
{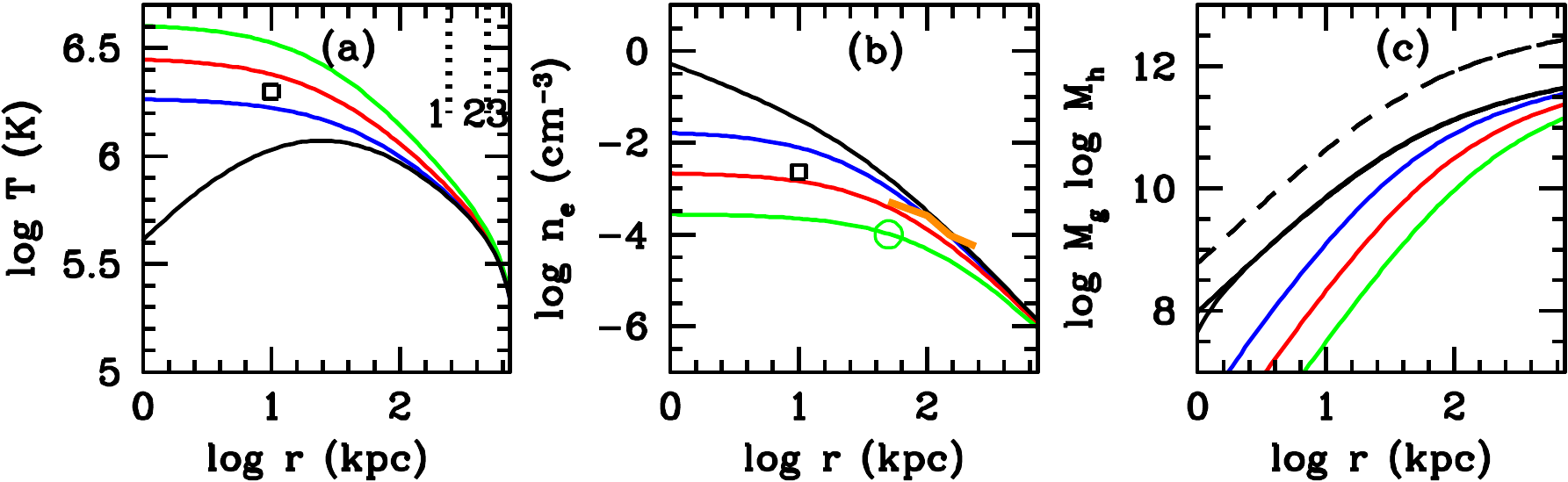}
\end{center}
\caption{
Four atmospheres in the $L^{\star}$ potential. 
{\it Black:} no-feedback atmosphere with 
reduced NFW profile,
{\it Blue, Red and Green:} three atmospheres with 
increasing feedback heating and distortion. 
{\it (a)} Temperature 
profiles. Radii at 1, 2 and 3 $r_{200}$ marked at 
upper right. 
{\it (b)} gas density profiles.
{\it Orange line}: $L^{\star}$ $n_e(r)$ profile 
for $50 < r < 240$ kpc from Oppenheimer et al. (2016).
{\it (c)} dark halo ({\it dashed line}) and 
cumulative gas 
mass profiles in $M_{\odot}$.
Black squares are Milky Way density and temperature 
at $r_{kpc} = 10$ from Miller \& Bregman (2016). 
}
\end{figure*}

In what follows we pursue the astrophysical implications 
of the observed OVI column density profile 
$N_{\rm OVI}(R)$, assuming that the oxygen has been expelled
from the central star-forming galaxies and now resides
in circumgalactic atmospheres in collisional ionization
equilibrium. 
Specifically,
we create simple hot gas atmospheres around a fiducial
$L^{\star}$ galaxy with a stellar mass
$M_{\star} = 10^{10.4}$ $M_{\odot}$
near the center of
the sample observed by 
Tumlinson et al. (2011), 
having a dark halo mass $M_{\rm h} = 10^{12.2}$ $M_{\odot}$.             
We begin with a reference atmosphere 
devoid of feedback distortion in which the gas has
an NFW profile reduced by the cosmic baryon fraction. 
From this we construct several approximate 
atmospheres distorted by increasing amounts of feedback 
heating which must be maintained by continued 
feedback that balances radiation losses.
The radial oxygen abundance profile in each atmosphere 
is adjusted until the space density profile 
of O$^{+5}$ ions is consistent with the observed
mean column density profile $N_{\rm OVI}(R)$. 
We show that
the local O abundance can significantly exceed solar 
at $\sim 100$ kpc with large total oxygen masses 
$\gta 10^9$ $M_{\odot}$. 
Radiative cooling time profiles are almost identical 
for all model atmospheres, but their radiative 
luminosities require feedback energies in excess
of that expected from supernova feedback alone.
This provides strong evidence that central black holes
are the dominant source of feedback energy 
having black hole accretion rates and masses 
that match those expected in $L^{\star}$ galaxies.

\section{Hydrostatic Circumgalactic Atmospheres}

Following 
Oppenheimer et al. (2016),
we adopt a representative $L^{\star}$ central galaxy
having stellar mass
$M_{\star} = 10^{10.4}$ $M_{\odot}$ 
and dark halo mass 
$M_{\rm h} = 10^{12.2}$ $M_{\odot}$.
The NFW halo that confines the hot gas is 
assumed to have concentration
$c = 4.86(M_{200}/10^{14}M_{\odot})^{-0.11} = 7.67$
(Neto et al. 2007),
and virial radius $r_{200} = 240$ kpc 
where the density is 1/200 of the critical density.
We expect that dark halos in $L^{\star}$ galaxies
are no longer 
accreting dark (or baryonic) matter 
and may have been quiescent for several Gyrs
(Prada et al. 2006; Cuesta et al. 2008;
Diemer \& Kravtsov 2014). 
Consequently, we adopt an NFW density profile
for the dark matter halo, 
and assume that it has not changed substantially 
since redshifts $z \sim 0.2$ where most
circumgalactic OVI absorption is observed.

We now construct several approximate atmospheres in 
this halo having increasing amounts 
of distortion due to feedback.
The equation of hydrostatic equilibrium 
$dP/dr = - \rho g_{NFW}$
can be written
\begin{equation}
dT/dr = - (T/ r)(d\log \rho / d \log r)
- g_{\rm NFW}(\mu m_p/ k)
\end{equation}
where $P = (k/\mu m_p)\rho T$,
$m_p$ is the proton mass and $\mu = 0.61$
is the molecular weight.
In the absence of feedback, 
baryons also have an NFW density profile reduced
by the cosmic baryon fraction, $f_b = 0.16$.
We disregard the stellar mass of the 
central galaxy since it is only $\sim$10 percent 
of the baryonic mass within the virial radius.
The gas density profile without feedback 
is therefore 
$\rho(r) = f_b \rho_0 /y(1+y)^2$ 
where $y = c(r/r_{200})$,
$\rho_0 = (200\rho_c/3)c^3/f(c)$,
$f(y) = \ln(1+y) - y/(1+y)$,
and $\rho_c = 9.2 \times 10^{-30}$ gm cm$^{-3}$
is the critical density for Hubble constant
$H_0 = 70$ km s$^{-1}$.
The halo mass is 
$M_h = M_{200} = (4\pi /3)200\rho_c r_{200}^3$
and the gravitational acceleration 
$g_{NFW} = GM_h(r)/r^2$.

In the presence of feedback,
atmospheric gas is heated,
its central entropy is increased  
and the entire atmosphere is pushed outward. 
To mimic these feedback effects,
we consider density profiles described by
\begin{equation}
\rho = f_b \rho_0/ (y_0 + y)^3
\end{equation}
where $y_0$ is a parameter that increases with 
the influence of feedback.
The central entropy $S = T/n_e^{2/3}$ 
has zero slope similar to galaxy group profiles 
visible in X-rays (Pratt et al. 2010).
Since feedback sources are centrally concentrated,
these atmospheres are designed to approach 
the feedback-free density profile at large radius, 
$\rho(y) \approx f_b \rho_0/y^3$.
We therefore disregard any significant density 
change at large radius, either 
an enhancement as gas is pushed out by feedback 
or a gas deficiency if feedback 
extends to the distant halo.
In any case, intermediate radii, $\sim 100$ kpc, 
are most relevant to our concerns here.
Our adoption of equation (2) does not affect 
conclusions discussed below.

Solid black contours in Figure 1 show temperature, density,
and gas mass profiles
for a circumgalactic atmosphere
without feedback.
Also in Figure 1 
are profiles for
three atmospheres with increasing feedback:
\begin{equation}
{\rm blue:red:green ::}~y_0 = 1:2:4
\end{equation}
Atmospheric structural distortion 
is stabilized when the time-averaged 
feedback power maintains 
atmospheric profiles against radiation losses.
If feedback drops below this maintenance level,
large masses of gas can
cool toward the central galaxy,
resulting in an unphysical mass accumulation,
similar to the overcooling problem
encountered in cosmological simulations.
Structural and maintenance feedback 
can be provided
by the same physical mechanisms.

The green atmosphere in Figure 1 is 
designed to match  
a hot gas density 
of $10^{-4}$ cm$^{-3}$ at radius 
$\sim$50-60 kpc in the Milky Way
as suggested by Fang et al. (2013) 
from XMM surface brightness observations
and by Sokolowska et al. (2016) 
from OVII and OVIII observations.
This density is marked with 
a green circle in Figure 1b.

\clearpage

\begin{figure*}[ht]
\minipage[t]{0.32\textwidth}

\begin{center}
\includegraphics[width=5.6cm]
{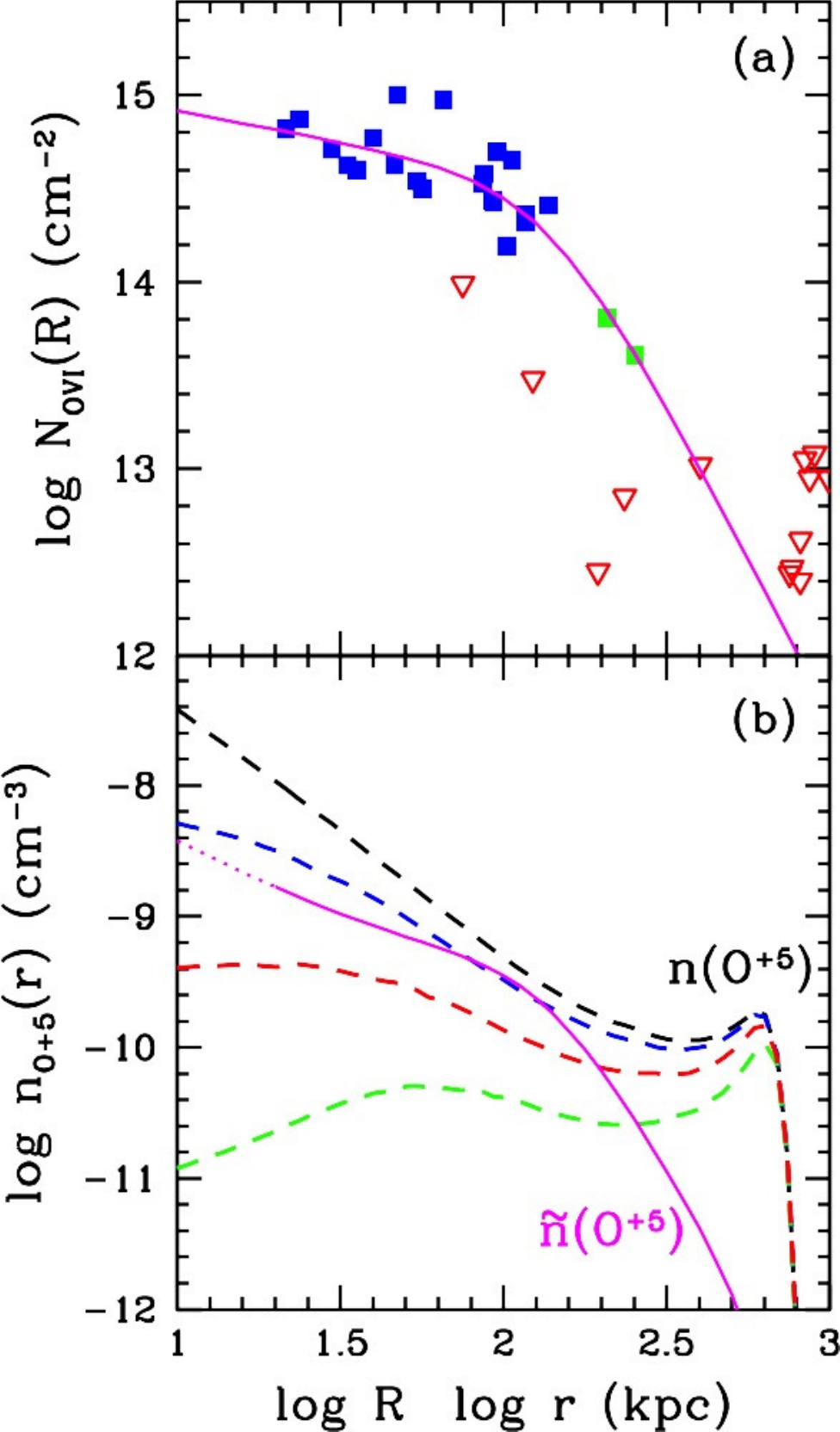}
\end{center}
\caption{
{\it (a)} N$_{\rm OVI}$ observations.
Blue squares from COS-Halos,
green squares and red
upper limits from JCM.
{\it (b)}
{\it Dashed lines:} $n_{O+5}(r)$ profiles for
the four atmospheres, all
with solar abundance.
{\it Magenta line:}
${\tilde{\rm n}}_{O+5}(r)$ profile. 
}

\endminipage\hfill
\minipage[t]{0.32\textwidth}

\begin{center}
\includegraphics[width=5.4cm]
{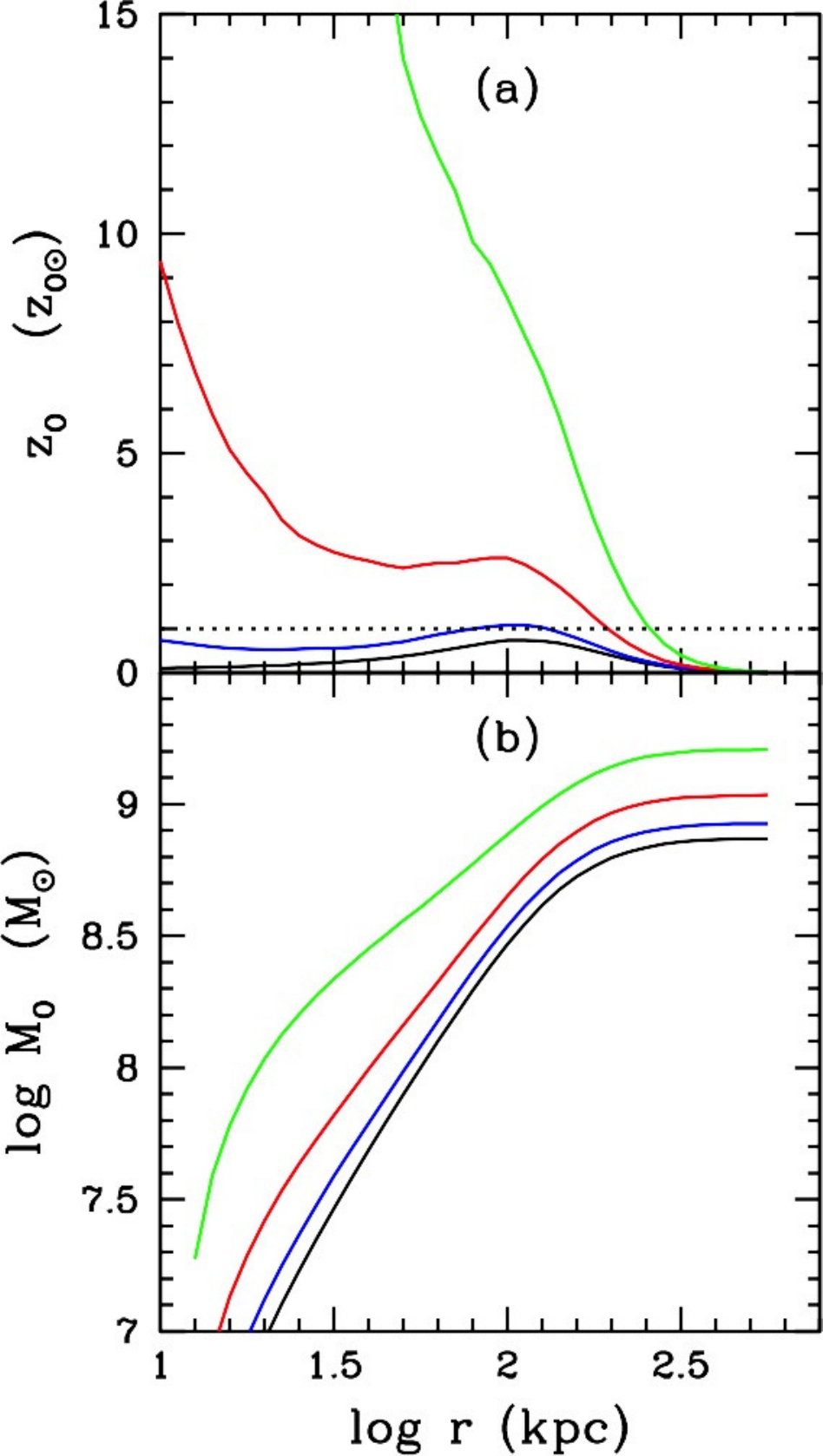}
\end{center}
\caption{
{\it (a)} 
Oxygen abundance profiles 
in solar units required 
for each atmosphere;
{\it Dotted line} marks solar abundance.
{\it (b)}
Cumulative oxygen mass profiles.
}

\endminipage\hfill
\minipage[t]{0.32\textwidth}%

\begin{center}
\includegraphics[width=5.3cm]
{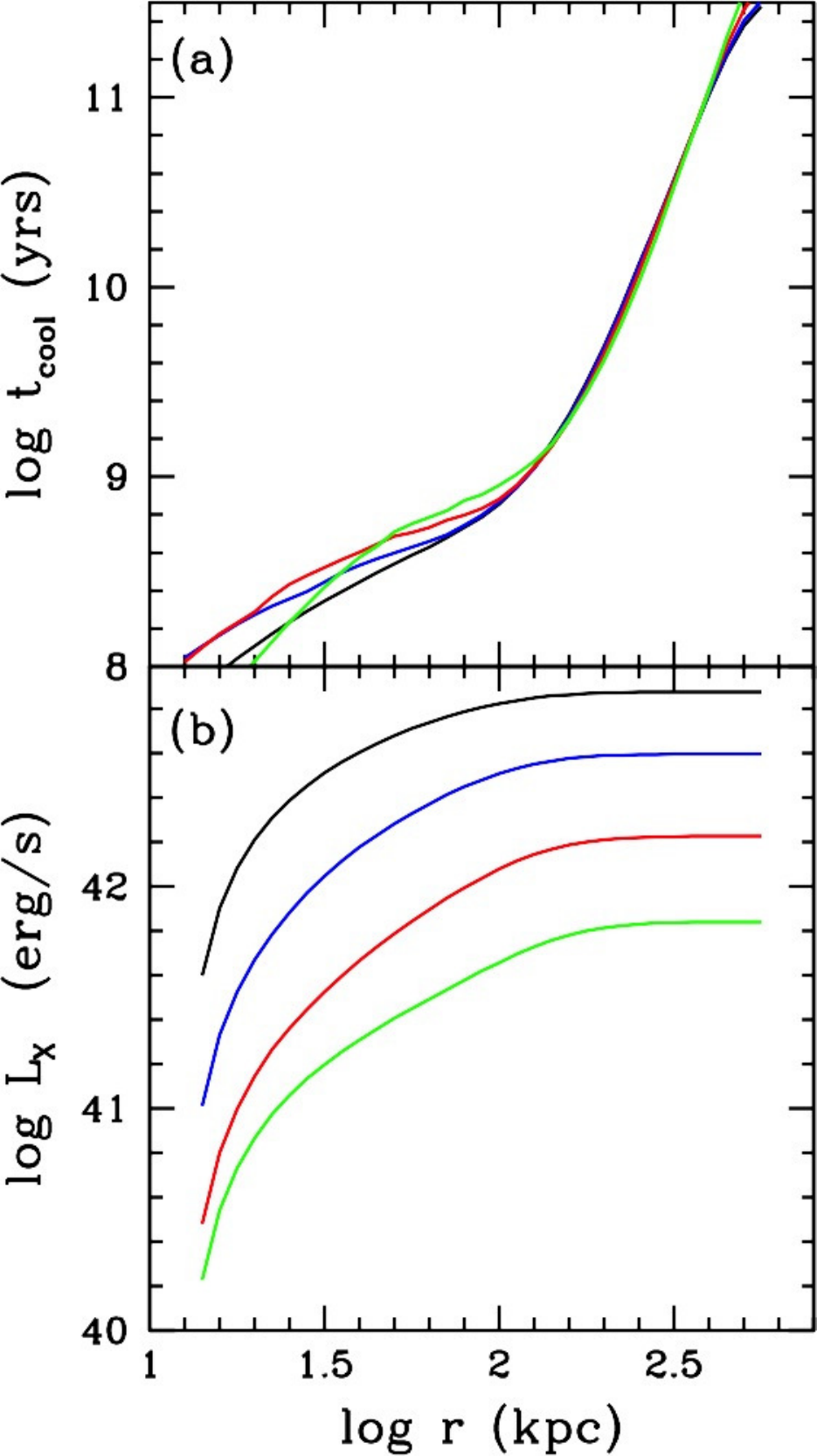}
\end{center}
\caption{
{\it (a)}
Short cooling time profiles for each atmosphere.
{\it (b)}
Cumulative bolometric luminosity profiles 
of each atmosphere, emitted primarily at
energies $<0.5$\,keV.
}

\endminipage
\end{figure*}

\section{O$^{+5}$ Column and Space Densities}

Blue squares in Figure 2a are 
COS-Halos detections of OVI column densities
$N_{\rm OVI}$ in circumgalactic atmospheres of 
$L^{\star}$ galaxies (Tumlinson et al. 2011)
at sightline offsets $R$ in kpc.
While most columns are clustered about 
$N_{\rm OVI} \sim 10^{14.7}$ cm$^{-2}$, 
there is evidence of decreasing columns 
beyond $R \sim 100$ kpc.
This decrease is confirmed by 
additional OVI observations by JCM, 
many having larger $R$.
In Fig. 2a we plot JCM data for
star-forming $L^{\star}$ galaxies 
having impact offsets $R > 75$ kpc, 
and stellar masses within 
10.1 $< {\rm log}(M_{\star}/M_{\odot}) <$ 10.7.
While only two galaxies are detected 
(green squares in Figure 2), 
a large number of upper 
limits (inverted red triangles) provides
convincing evidence of a rather 
sharp decrease in OVI columns 
beyond $\sim 100$ kpc.
Such a decrease strongly 
supports the prevailing interpretation
that very large oxygen masses 
have been ejected from $L^{\star}$ galaxies 
and/or their progenitors.
Decreasing $N_{\rm OVI}(R)$ profiles also 
constrain the atmospheric O/H abundance
and the feedback mechanism that 
maintains it.

Although data are sparse, we fashion
a magenta line in Figure 2 showing a likely 
column density profile for $L^{\star}$ galaxies.
The magenta line is described by
\begin{equation}
N_{\rm OVI} = N_0 [1 + (R/R_0)^p]^{-1} (R/R_1)^{q} 
\end{equation}
where 
$N_0 = 5.2 \times 10^{14}$ cm$^{-2}$,
$R_1 = 40$ kpc,
$q = -1/3$, $R_0 = 140$ kpc and $p = 3$.

The corresponding empirical space
density of O$^{+5}$ ions $\tilde{n}_{O+5}(r)$,
can be found from an Abel integral inversion,
\begin{displaymath}
{\rm \tilde{n}}_{O+5}(r) = - {1 \over \pi}
\int_r^{\infty} {d N\over dR} {dR \over (R^2 - r^2)^{1/2}}
\end{displaymath}
which is plotted
as a magenta line in Figure 2b.
The inner, dotted part of the 
${\tilde{\rm n}}_{5} \equiv 
{\tilde{\rm n}}_{O+5}(r)$ profile
is unconstrained by current N$_{\rm OVI}$ data.
For comparison, dashed lines in Figure 2b shows  
O$^{+5}$ space density profiles 
$n_5(r)$
for each atmosphere in Fig. 1,
assuming uniform solar abundance 
$A_{O\odot}= 5 \times 10^{-4}$ and 
employing ionic fractions
for collisional ionization equilibrium 
(Gnat and Sternberg 2007).

The juxtaposition of $n_5$ and $\tilde{n}_5$ profiles
in Figure 2b allows an instant determination of 
atmospheric O abundance profiles $Z_O(r)$ 
in solar units required to 
match the mean profile 
$\tilde{n}_5(r)$ and, when projected,
the adopted 
mean column density profile $N_{\rm OVI}(R)$
in Figure 2a.
For example, the oxygen abundance in solar units
for the red atmosphere is simply 
$Z_{O,red}(r) = {\rm \tilde{n}}_5(r)
/n_{5,red}(r)$ 
at every radius.

Oxygen abundance profiles derived in this manner
for each atmosphere are illustrated in Figure 3a.
Maximum abundances are expected 
near $r \sim 100$ kpc 
where the slope $dN_{\rm OVI}/dR$ changes.
Oxygen abundances $Z_O$
are large, particularly in the high feedback, 
low density green atmosphere.
At $r = 100$ kpc, O abundances in the
blue, red and green atmospheres 
are $ Z_O = $1.1, 2.6 and 8.5 solar respectively.
The cumulative oxygen mass shown in Figure 3b
also increases with feedback in the 
blue, red and green atmospheres, totaling
0.84, 1.08 and 1.61 (in units of $10^9$ $M_{\odot}$)
respectively.
In a closed box model
an $M_{\star} = 10^{10.4}$ $M_{\odot}$
stellar mass can produce an oxygen mass
of only $\sim10^8$ $M_{\odot}$ (Zahid et al. 2012).
However, in the simulation of Oppenheimer
et al. (2016) the total oxygen mass within
$r_{200}$ at zero redshift
is $3 \times 10^8$ $M_{\odot}$, 
but their computed sightline
column densities $N_{\rm OVI}$ are 
about 3 times lower than those observed 
and the space density of O ions is not as 
concentrated near $r \sim 100$ kpc 
as we require here.


In Figure 4a we show the remarkable
similarity of cooling time profiles
$t_{cool}(r)$ for all atmospheres.
It is easy to show that this similarity 
follows from the relation
$Z_O = \tilde{{\rm n}_5}/n_{5}$
when oxygen dominates the metallicity,
$Z \approx Z_O$.
Therefore, the cooling time profile 
holds for any $L^{\star}$ atmosphere,
not just those proposed here.

Large bolometric X-ray luminosity profiles
in Figure 4b, 
$\log L_X \approx 42.2\pm0.4$ erg s$^{-1}$,
attest to the 
powerful cooling effect of $\sim 10^9$ $M_{\odot}$
of oxygen extending to $r \approx 100$ kpc.
To explore the destiny of pure cooling without
feedback, we computed spherical time-dependent 
cooling flows allowing
blue, red and green (b,r,g) atmospheres 
initially at rest 
to evolve for the lookback
time 2.4 Gyrs at redshift 0.2 
with abundances from Fig. 3a. 
These straightforward 
calculations, not discussed in detail here,
verify that gas masses of $\log(M/M_{\odot}) 
\approx (11.2$, 10.9 and 10.5), 
all in excess of $M_*$,
cool at the origin in (b,r,g) atmospheres 
during this relatively short time.
All oxygen-rich gas within $\approx 210$\,kpc cools.
Such an atmospheric collapse is not supported by
the recent star formation history
of the Milky Way (Gonzalez et al. 2017)
or by observations of 
extended circumgalactic OVI 
absorption in star-forming $L^{\star}$ 
galaxies.
Clearly, a strong
time-averaged  maintenance feedback
power comparable to $L_X$
must be provided by supernovae or 
central black holes.
Our concern here is not the energetics of 
the outflows that previously carried O-rich gas out to
$\sim 100$ kpc, but how this gas  
is maintained during more recent times.

In the figures we assume that oxygen is
completely mixed on atomic scales, 
but the degree of mixing remains uncertain.
It is likely that circumgalactic oxygen
is inhomogeneous, which would decrease the
radiative cooling time
below that in Figure 4a,
causing small, O-rich regions to cool locally,
perhaps even in the presence of maintenance feedback.
Such local cooling is supported by low-ionization
circumgalactic absorption lines in quasar spectra.

Core-collapse supernovae in 
late type $L^{\star}$ galaxies 
occur at a rate of 1-3 per century, but generate only
$\sim 0.6 \times 10^{42}$ erg s$^{-1}$,
most of which is dissipated locally in strong
shocks in the galactic disk.
Successful $L^{\star}$ feedback requires 
intermittent shocks propagating through gas 
at $r \sim 100$ kpc at some mean interval
$\Delta t$.
The rate that entropy is radiated
away by gas at $r = 100$ kpc 
(where $n \approx 10^{-4}$ cm$^{-3}$,
$T \approx 10^6 $ K and $Z \approx Z_{\odot}$)
can be restored by shocks every 
$\log\Delta t =$ (7, 8, 9) yrs 
with large Mach numbers (1.25, 1.7, 3.9),
roughly similar to feedback shock strengths 
in galaxy group atmospheres.
Even if supernova power were sufficient, 
neither stochastic variations in the 
galactic supernova rate
nor slowly moving (forward and reverse) shocks
expected in sustained supernova-driven winds 
are sufficiently intermittent.

Consequently, massive central black holes may  
be the most promising source of 
distant, intermittent feedback shocks.
Following King and Pounds (2015), 
fast-wind feedback from 
AGNs having luminous accretion disks
generate mechanical 
luminosities $\sim 0.05L_E$ 
where $L_{E} = 1.26 \times 10^{38}
(M_{\bullet}/M_{\odot})$ erg s$^{-1}$
is the Eddington luminosity. 
Since central black holes in 
$L^{\star}$ galaxies with 
$M_{\star} = 10^{10.4}$ $M_{\odot}$ 
have masses $M_{\bullet}$ 
similar to that in the Milky Way,
$4.5\times 10^6$ $M_{\odot}$,
the expected mechanical power of $L^{\star}$ winds is
$\sim 30 \times 10^{42} \sim 30 L_X$ erg s$^{-1}$. 
This is sufficient to balance X-ray luminosities in 
Figure 4b with central AGNs that  
are intermittently active only 3 percent of the time. 
This duty cycle is  
$\lta$10 percent as required
by Oppenheimer et al. (2017) 
to enhance OVI-absorbing ions in 
atmospheres of $L^{\star}$ galaxies.

More collimated feedback modes 
may also be present.
In galaxy groups, having
black holes and dark halo masses roughly
ten times larger than in $L^{\star}$ galaxies,
the accretion power of central black holes
maintain atmospheric
temperatures $T \sim 10^7$ K with no evidence
of recent star formation or
luminous accretion disks.
In addition, the Fermi bubbles
in the Milky Way reveal that its central
black hole is currently delivering an
(albeit uncertain) energy of $\sim 10^{59}$
ergs to the circum-Galactic atmosphere
(Guo \& Mathews 2012, Guo et al. 2012),
equivalent to the total Galactic supernova energy
generated in $10^8$ yrs.
Since outward propagaing shocks in 
atmospheres with $d\log \rho / d \log r > -2$ 
naturally weaken, shocks formed 
at large radii by collimated 
black hole outflows provide more efficient feedback.
The non-thermal content of the Fermi Bubbles
and extragalactic jets 
may also suggest feedback modes stronger than 
disk winds.


Further evidence that central black holes
may energize circumgalactic gas are 
much lower OVI absorption columns observed in
non-starforming passive $L^{\star}$ galaxies
(Tumlinson et al. 2011).
Since central black holes in passive $L^{\star}$ galaxies 
are on average $\sim 40$ times more massive 
than those in star-forming $L^{\star}$ galaxies
(Reines \& Volonteri 2015),
the accretion feedback in passive galaxies 
may drive most of the OVI-absorbing 
circumgalactic gas entirely 
out of the galactic potential.
Upper limits of four star-forming
$L^{\star}$ galaxies in Figure 2a 
also lie far below the magenta profile.
Perhaps unusually large feedback events
with energy $\gta (G M_{g}M_h/r)|_{100 kpc} \sim 10^{58.7}$ 
ergs have removed the entire atmosphere.

\vspace{0.3cm}

We acknowledge a very helpful observational communication 
from Todd Tripp and useful remarks from Fabrizio Brighenti.
WGM thanks George Poultsides, John Harris, Joseph Palascak 
and Francisco Rhein for sustaining support.

\end{document}